\documentstyle[aps,prb,twocolumn]{revtex}

\begin{document}
\draft
\title{Berry's phase for large spins in external fields}

\author{V. A. Kalatsky$^{\rm a}$, E. M\"uller--Hartmann$^{\rm b}$,
V. L. Pokrovsky$^{\rm a, b, c}$, G. S. Uhrig$^{\rm b}$}

\address{$^{\rm a}$Department of Physics, Texas A\&{}M University,
 College Station, Texas 77843-4242}
\address{$^{\rm b}$Institut f\"ur Theoretische Physik,
 Universit\"at zu K\"oln,  D-50937 K\"oln}
\address{$^{\rm c}$Landau Institute, Chernogolovka}

\date{\today}

\maketitle

\begin{abstract}
It is shown that even for large spins $J$ the fundamental difference
between integer and half-integer spins persists. In a quasi-classical
description this difference enters via Berry's connection.
This general phenomenon is derived and illustrated for large spins confined
 to a plane by crystalline electric fields. Physical realizations are
rare-earth Nickel Borocarbides. Magnetic moments for half-integer spin
 (Dy$^{3+}$, $J=15/2$) and 
magnetic  susceptibilities for integer spin (Ho$^{3+}$, $J=8$) are calculated.
Experiments are proposed to furnish evidence for the predicted
fundamental difference.
\end{abstract}

The common belief that large spins are equivalent to classical
 charged gyroscopes
was strikingly undermined by Haldane \cite{haldane} who indicated that the
groundstate and low-lying states of magnetic chains consisting of localized 
spins $J$ are different for integer
and half-integer $J$ even if $J$ is large. Here we consider
a similar effect for individual large spins placed in an external fields such
as the crystalline electric field (CEF) and the external magnetic field. 
Its origin can be traced back to a geometric phase occurring on
passing from quantum mechanical to quasi-classical spins.

The general Hamitonian of a localized spin associated with a magnetic 
moment can be written as a function
of its components:
\begin{equation}
H_S\,=\,\hat{f}(\vec{J})-\vec{h}\vec{J}
\label{hamilton}
\end{equation}
where $\hat{f}(\vec{J})$ is an arbitrary function of $\vec{J}$ 
satisfying only two requirements: it is a Hermitian operator and it is 
an even function
of $\vec{J}$: $\hat{f}(\vec{J})=\hat{f}(-\vec{J})$.
The latter requirement is equivalent to time-reversal symmetry
\cite{newma71}.
In this paper we will focus on localized moments which are
essentially confined to a plane. The confinement is due to
the effects of the crystalline electric field.
The important degree of  freedom is rotation about the  normal
vector of a certain plane which we choose as $z$-direction. It is then
appropriate to introduce the azimuthal angle $\varphi$  and its 
conjugate momentum  $J_z$ as canonical variables. Setting $\hbar = 1$ we
may use
\begin{mathletters}
\label{J-phi}
\begin{eqnarray}
J_z &=& -i\frac{\partial}{\partial \varphi}
\label{J-phia}
\\
 J_x &=& \sqrt{J(J+1)-J_z^2}\cos\varphi
\\
J_y &=& \sqrt{J(J+1)-J_z^2}\sin\varphi \ .
\end{eqnarray}
\end{mathletters} \noindent
Eqs. (\ref{J-phi}) generally have a symbolical meaning since the operators 
on the right hand sides should be ordered to guarantee that 
${J_x^+} =J_x;\;J_y^+=J_y$ and
that the canonical permutation relations are satisfied. However, at large $J$ 
the non-commutativity is  small and we do not need to bother with
the symmetrization. Inserting
(\ref{J-phi}) into the Hamitonian (\ref{hamilton}),  the 
problem is reduced to the 
solution of a Schr\"{o}dinger wave equation with, in general, a complicated 
differential operator.

So far, no difference between integer and half-integer spins occured. 
The difference resides in a different global phase behavior.
For large spins $J$ one may pass to a  quasi-classical
description via the overcomplete set of coherent states $\psi_{\vec{n}}$
with 
\begin{equation}
\label{defpsi}
\langle \psi_{\vec{n}} | \vec{J} |\psi_{\vec{n}}\rangle = J\vec{n}
\ .
\end{equation}
Moving the spin induces a certain orbit of the tips of the 
unit vectors $\vec{n}$  on the unit sphere $S^2$. If this orbit is
closed there is {\em no} difference in the purely classical picture  between 
starting point and end point. Due to Berry's connection
\cite{berry84}, however, this is not the whole story. In order to be 
quantitative we first have to fix the phases of the $\psi_{\vec{n}}$ in
(\ref{defpsi}). The natural choice is to fix the phase of $\psi_{\vec{z}}$
and to take \cite{moody86a}
\begin{equation}
\psi_{\vec{n}} = \exp(iJ_z\varphi)  \exp(iJ_x\theta)  \exp(-iJ_z\varphi)
 \psi_{\vec{z}}
\end{equation}
where $(\varphi,\theta)$ are the angles characterizing the unit vector
$\vec{n}$ ($\vec{z}$ being the unit vector in $z$-direction).
Berry's connection is given for the single, non-degenerate state
$\psi_{\vec{n}}$ (abelian case\cite{berry84,wilcz84,shape89}) by
$\vec{A} = \langle \psi_{\vec{n}} |i\vec{\nabla}| \psi_{\vec{n}}\rangle$.
Since $\vec{n}$ is confined to the unit sphere, spherical coordinates are
most appropriate and only the $A_\theta$ and the $A_\varphi$ components matter.
One finds
\begin{mathletters}
\begin{eqnarray}
A_\theta &=& \langle \psi_{\vec{n}} | i\partial/(\partial\theta)
| \psi_{\vec{n}} \rangle 
 = -  \langle \psi_{\vec{z}} | J_x | \psi_{\vec{z}} \rangle
= 0 \\  \nonumber
A_\varphi &=&\langle \psi_{\vec{n}} | i \sin(\theta)^{-1}
\partial/(\partial\varphi)
| \psi_{\vec{n}} \rangle \\ \nonumber
 &=& \sin(\theta)^{-1} \langle \psi_{\vec{z}} |J_z(1-\cos(\theta))
+\sin(\theta)J_y | \psi_{\vec{z}} \rangle \\
&=& J(1-\cos(\theta))/\sin(\theta) \ .
\end{eqnarray}
\end{mathletters}
This connection then gives rise to the geometrical phase
$\exp\big(i\int_\gamma \vec{A} d\vec{l}\big)$ along the path $\gamma$. 
Since we are interested in this work in  motion in the $xy$-plane only, we
have $\theta=\pi/2$ and thus the phase is
$\exp(iJ\Delta\varphi)$. As was to be expected from the physical 
origin of this phase it can  be put to zero {\em locally} by an appropriate
gauge. Globally, i.e.\ for complete tours of $\Delta\varphi=2\pi$, one
sees that no effect occurs for integer $J$ but that for half-integer $J$ a
factor -1 applies which cannot be gauged away. One way to account for
the phase behavior is to use (\ref{J-phia}) and {\it anti}periodic boundary
condition for half-integer spins and periodic boundary
condition for integer spins, respectively. A more elegant way is
to stick to the connection $\vec{A}$. This means we use
\begin{equation}
\label{J-phia-neu}
J_z = -i\left( \frac{\partial}{\partial \varphi} - i A_\varphi \right)
\end{equation}
instead of (\ref{J-phia}). Since, however, a change of gauge can
alter $A_\varphi$ by any integer value one may use (\ref{J-phia})
for integer spins. For half-integer spins we use 
$J_z = -i\left( \partial/(\partial \varphi) - i/2 \right)$.

 Thus, despite of the large value of $J$, the low-lying 
states of the Hamiltonian (\ref{hamilton}) for
 integer and half-integer spins are fundamentally different.
 In the absence of an external
magnetic field all stationary states of a half-integer spin are doubly
degenerate (Kramers degeneracy \cite{abrag}).
This means that, in analogy to the linear Stark effect,
 the groundstate may be characterized by a finite magnetic moment.
 To the contrary,  the groundstate of an integer spin is non-degenerate
for sufficiently low crystalline symmetry.
 Therefore, the magnetization in the groundstate is exactly
zero and  depends linearly on the magnetic field as long as the field is weak
enough.

To be more specific, we consider an important application of these general
ideas to the ions of the rare earth elements Ho$^{3+}$ and Dy$^{3+}$.
The triply charged ions are decisive for the magnetic moments in the compounds 
$R$Ni$_2$B$_2$C ($R$ stands for a rare earth element) the properties of which
attracted much attention in the last few years
\cite{discovery,siegrist,canfield,naugle}. The
ion Ho$^{3+}$ has 10 electrons in the 4f-shell. According to Hund's rule, 
it has the total spin $S =2$, the orbital moment $L =6$ and the total moment 
$J =8$. The corresponding numbers for Dy$^{3+}$ (9 electrons in the 4f-shell)
 are: $S =\frac{5}{2}$, $L =5$, $J =\frac{15}{2}$. Thus, both values of $J$ 
are rather large and close to each other.

All the compounds ($R$ =Y, La, Ho, Dy, Tm, Tb, Er, Yb) crystallize
 as perovskites with the rare earth ions forming a tetragonal centered lattice
\cite{siegrist}. The magnetic moments of the Ho and Dy compounds are confined
presumably in the $ab$-plane thus realizing the situation discussed above.
The simplest crystal electric field (CEF) 
Hamitonian $H_{\rm CEF}$ displaying tetragonal
symmetry reads:
\begin{equation}
H_{\rm CEF}\,=\,\frac{a}{2}J_z^2 - 2b(J_x^4+J_y^4).
\label{CEF}
\end{equation}
with $a,b >0$. Quartic (and higher) terms in $J_z$ are neglected since we
assume $J_z\ll J_x, J_y$. For the same reason one can introduce the coordinate 
$\varphi$ in a slightly simplified way (compared to (\ref{J-phi})):
 $J_x =J\cos\varphi$, $J_y =J\sin\varphi$, and 
$J_z =-i\partial /(\partial\varphi)+A_\varphi$ with 
$A_\varphi=0$ for integer J and  $A_\varphi=1/2$ for half-integer J.
 Together with the magnetic field 
contribution, the total Hamiltonian becomes:
\begin{eqnarray}\nonumber
H &=& -\frac{a}{2}\left(
\frac{\partial}{\partial\varphi}-iA_\varphi\right)^2  
-  \frac{b}{2}J^4(3+\cos(4\varphi)) \\
&&- h J\cos(\varphi-\varphi_h),
\label{H-phi}
\end{eqnarray}
where $\varphi_h$ is the angle the magnetic field forms with the $x$-axis. 
The effective field $h$ is the magnetic field times $g\mu_{\rm B}$.
The Hamiltonian (\ref{H-phi}) was found in a previous work \cite{KP} in which 
it was also analyzed for integer $J$.

It will be shown later that the CEF constants $a$ and $bJ^2$ are of the same 
order of magnitude. Due to the
large value of $J$ the potential energy has very deep minima near the points
$\varphi = \varphi_l = l\pi/2, l\in\{0,1,2,3\} $. Let us denote the 
oscillatory states 
localized near each value of $\varphi_l$ as $|l\rangle$. 
To be more specific, each $|l\rangle$ is the groundstate
\cite{note1} of
$H_l=-(a/2)\partial^2/(\partial\varphi^2)+ U_l(\varphi)$ with
\begin{equation} 
\label{udef}
U_l(\varphi) = \left\{
\begin{array}{cc}
 -  \frac{b J^4}{2}(3+\cos(4\varphi)) & {\rm for} \
|\varphi-l\frac{\pi}{2}| \le \frac{\pi}{4}  \\
2bJ^4 & {\rm otherwise}
\end{array} \right. .
\end{equation}
 Without loss of generality
we assume that the corresponding eigenenergy is zero. 
 Neglecting the 
overlap between different $|l\rangle$, we find that the energy level is 
fourfold degenerate. The overlap lifts partly the degeneracy even in the
absence of magnetic field. In complete analogy to the derivation of
the dispersion of a tight-binding model \cite{ashcr76} we define the
hopping matrix element
\begin{equation}
\label{wdef}
w = -\langle l| H - H_{l+1}|l+1\rangle > 0\ .
\end{equation}
The hopping part of effective Hamiltonian is
\begin{equation} 
\label{Hdef}
H_w = -w(C(\alpha)+C^+(\alpha))
\end{equation}
 where $C$ is the unitary rotation operator
which induces $|l\rangle \to |l+1\rangle \exp(i\alpha)$
and $|3\rangle \to |0\rangle \exp(i\alpha)$.
For integer spin we have $\alpha=0$, for half-integer spin we use
$\alpha=\pi/4$ resulting from 
\begin{equation}
\label{alphadef}
\exp(i\alpha) = 
\exp\Big(i\int_{l\pi/2}^{(l+1)\pi/2} A_\varphi d\varphi\Big)\ .
\end{equation}
 This is the direct effect of the connection $A_\varphi$
in absolute analogy to the Peierls phase in tight-binding models
in magnetic fields.
For what follows it is important to note that $w$ is exponentially
small if the wells at $\varphi_l$ are deep enough such that the
ground states $|l\rangle$ are well, i.e.\ exponentially, localized.
To estimate $w$ we use the ansatz $|l\rangle
\propto \exp(-| \int_{l\pi/2}^\varphi 
\sqrt{2(U_l(\varphi)-U_{\rm min})/a} d\varphi|) $
which is motivated by the groundstate for the harmonic potential
close to the minima and its natural extension in a WKB-type approach.
The main contribution to the $\varphi$-integral for $w$ comes
from the vicinity of $\varphi = \pi/4$ for $l=0$ and leads to
\begin{equation}
\label{wcalcul2}
w \propto \exp\left(-\sqrt{\frac{2bJ^4}{a}}\right)\ .
\end{equation}

The eigenstates and eigenvalues of (\ref{Hdef}) are easily read off
since we deal with a translationally invariant, $d=1$, four-site 
tight-binding model. Thus the eigenstates are characterized by
some momentum $k\in\{ 0,\pm\pi/2,\pi \}$
\begin{equation}
\label{eigenstates}
\psi_k = \frac{1}{2} \sum\limits_{l=0}^3 \exp(ikl) |l\rangle \ .
\end{equation}
The corresponding eigenenergies are
\begin{equation}
\label{eigenvalues}
E_k = -2w\cos(k+\alpha)\ .
\end{equation}
Thus for integer $J$ ($\alpha=0$) there is a non-degenerate groundstate
$(k=0,E_k=-2w)$, two degenerate excited states
$(k=\pm\pi/2,E_k=0)$, and the highest excited state 
$(k=\pi,E_k=2w)$. For half-integer $J$ ($\alpha=\pi/4$)
both the groundstate and the excited state are doubly degenerate
with $(k=0, -\pi/2,E_k=-\sqrt{2}w)$ and 
$(k= \pi/2, \pi,E_k=\sqrt{2}w)$.

The main difference between integer and half-integer spin resides here
in a different degeneracy of the ground states. Physically this
difference becomes manifest, for instance, when a magnetic field
$h$ is applied. We first consider an in-plane field as in
(\ref{H-phi}). For well-localized states $|l\rangle$ 
such a magnetic field is site-diagonal with matrix elements
\begin{equation}
H_h = h J\cos(l\pi/4-\varphi_h)\ .
\end{equation}
The eigenvalues of $H_w+H_h$ can be given analytically. For
integer spins and half-integer spins one finds, respectively
\begin{mathletters}
\label{energies}
\begin{eqnarray}
\label{intJ}
E_{\rm int} &=& \pm\sqrt{2w^2+{\bar h^2\over 2}\pm
\sqrt{\left( 2w^2+{\bar h^2 \over 2} \right)^2-\bar h_x^2\bar h_y^2}}
\\
\label{halfJ}
E_{\rm hint} &=& \pm\sqrt{2w^2+{\bar h^2\over 2}\pm \bar h
\sqrt{ 2w^2+{\bar h^2\over 4}
-\frac{\bar h_x^2\bar h_y^2}{\bar h^2}}}.
\end{eqnarray}
\end{mathletters}
where we used $\bar h$ as a shorthand for $hJ$. 
From (\ref{energies}) one obtains for the groundstate energies in the limit
of small magnetic fields
\begin{mathletters}
\label{groundenergies}
\begin{eqnarray}
E_{\rm int} &\approx& -2w - h^2J^2/(4w)
\\
E_{\rm hint} &\approx& -\sqrt{2}w - hJ/2 .
\ .
\end{eqnarray}
\end{mathletters}
As expected the correction is quadratic in the non-degenerate,
integer $J$ case, but linear in the degenerate, half-integer $J$ case
(as in the linear Stark-effect).
So we have for integer $J$ a finite $T=0$ susceptibility
\begin{equation}
\label{chi}
\chi_{\rm plane} = -g^2\mu_{\rm B}^2(\partial E)^2/(\partial h^2)= 
g^2\mu_{\rm B}^2J^2/(2w) \ .
\end{equation} 
For half-integer $J$ the system shows a finite magnetic momentum
\begin{equation}
\label{moment}
\mu= -g\mu_{\rm B} \partial E/(\partial h) = g\mu_{\rm B} J/2
\end{equation}
 leading to a Curie
susceptibility diverging for $T\to0$. This constitutes an essential difference
between integer and half-integer spin. For large magnetic fields, however,
the difference vanishes since $w$ becomes unimportant.  Asymptotically
one obtains for
$hJ\gg w$
\begin{equation}
E \approx \pm hJ\cos(\varphi_h);\;\pm hJ\sin(\varphi_h)
\label{asympt}
\end{equation}
irrespective of whether $J$ is integer or half-integer.
This equation implies that the corresponding magnetic moments at
saturation are directed along one of the four easy axes. This intermediate
asymptotics is valid in the range $w \ll hJ \ll bJ^4$ where our 
tight-binding treatment stays valid. At significantly larger $h$ the
saturation magnetic moment is directed parallel to the magnetic field.
Comparing (\ref{moment}) and (\ref{asympt}) we, find that the finite
magnetic moment at low magnetic field for half-integer $J$ is 
precisely half of its saturation value ($9.8\mu_{\rm B}$ for
Dy$^{3+}$; cf.\ $10\mu_{\rm B}$ for
Ho$^{3+}$ with integer $J$).

Let us consider now the action of magnetic field along 
the  $z$-axis. To do so we rewrite 
\begin{eqnarray}
\nonumber
\frac{a}{2} J_z^2 - h J_z &=& \frac{a}{2}
\left(J_z-\frac{h}{a}\right)^2 - \frac{h^2}{2a} \\
&=& \frac{a}{2}\left(-\frac{i\partial}{\partial\varphi}+
A_\varphi-\frac{h}{a}\right)^2 - \frac{h^2}{2a} \ .\label{rewrite}
\end{eqnarray}
>From this we infer that the $z$-axis magnetic field adds a constant to
 the Hamiltonian and acts
as if the connection $A_\varphi$ is changed in the manner
$A'_\varphi = A_\varphi - h/a$.
Inserting $A'_\varphi$ in (\ref{alphadef}) shows that
 the Peierls phase 
$\alpha$ in (\ref{Hdef}) is changed like $\alpha \to 
\alpha - \pi h/(2a)$. The effect  of the Peierls phase change on 
the eigenenergies and on the groundstate energy in particular
is easily found in (\ref{eigenvalues}).
So, for integer $J$, we have $E=-2w\cos(\pi h/(2a)) - h^2/(2a)$ and 
from this $\chi_{\rm axis} = a^{-1} -\pi^2 w/(2a^2)$. Due to the 
exponential smallness of $w$ (see (\ref{wcalcul2})) the second term
is negligible compared to the first one and we find 
$\chi_{\rm axis} = a^{-1}$ for integer $J$.
Note that $\chi_{\rm axis}$ is exponentially small compared to
$\chi_{\rm plane}$ in (\ref{chi}) which proves the consistency
of our treatment for which we assumed that the moments are
essentially confined to the plane. For half-integer spin
we consider $E=-2w\cos(\pi h/(2a)-\pi/4) - h^2/(2a)$ and derive
$\mu = -g\mu_{\rm B}\partial E/(\partial h)= g\mu_{\rm B}w/(\sqrt{2}a)$ 
for the local moment in $z$-direction.
We obtain again a moment in $z$-direction which is exponentially small compared
to the one in the plane (\ref{moment}) consistent with the outset
of our theory. We observe that the difference between integer
and half-integer $J$ is also visible in the magnetic properties
perpendicular to the easy plane.

The quasi-classical treatment implying the splitting of each oscillatory
level into a quadruplet is valid provided that the level spacing
 between the localized oscillatory levels
$\omega$ is much larger than $w$. It is certainly 
incorrect for energies $E$ 
close to the maximum potential energy $U_{\rm max}=2bJ^2$.
One should have a rough idea on how many quadruplets our treatment
can be expected to apply to.
In the cases of interest $J=15/2, 8$ the total number of states is 16 or 17. 
Thus one can expect that one or two quadruplets are well described by the
effective four-site tight-binding model. 
This conclusion is confirmed by direct diagonalization of the 
17x17 matrix in a model crystal field 
\cite {harmon}.
In order to have an estimate for the ratio $b/a$ we compare the total
number of states $2J+1$ with the number of levels $N$ in the 
four wells of the
potential $U(\varphi )=-(bJ^4/2)(3+\cos(4\varphi))$ in the quasi-classical
approach
\begin{equation}
N = \frac{8}{\pi\hbar}\int_0^{\pi/4}\sqrt{2m(U_{\rm max}-U(\varphi))}d\varphi
= \sqrt{\frac{32bJ^4}{\pi^2a}}\ .
\end{equation}
The area in phase space of the maximum classical orbit is divided by 
$2\pi\hbar$ to obtain an estimate for the number of states.
Here $\hbar^2/m$ is set to $a$.
Equating $N$ to $2J+1\approx2J$, we find $bJ^2/a=\pi^2/8\approx 1.23$, i.e.\
they are of the same order of magnitude.

Let us discuss the experimental consequences of the difference between
integer and half-integer $J$ described above.
The  clearest manifestations would occur in dilute alloys of Ho and Dy in the
compounds Lu$_{1-x}$Ho$_x$Ni$_2$B$_2$C,
Lu$_{1-x}$Dy$_x$Ni$_2$B$_2$C or in similar alloys with Y
in place of Lu. Both Lu$^{3+}$ and Y$^{3+}$ have zero magnetic moment.
The direct check of our prediction would be the measurement of EPR spectra for 
these alloys at low temperature $T < w$ (at about 2K according to the estimate 
in \cite{harmon}). We expect that the lowest four levels are split as 
described above into singlet, doublet and singlet for Ho. 
The resonance frequency will not
depend much on the magnetic field as long as  $hJ\ll w$.
In the case of Dy-alloys, however, the EPR resonance frequency is
proportional to the magnetic field as long as this is small
due to the finite magnetic moment. One can also check the
nonlinear dependence of energies (\ref{intJ}, \ref{halfJ})
 on the magnetic field.

Another option is to measure the magnetic field at  a nucleus by NMR.
 The magnetic field at a nucleus must be essentially zero for Ho-alloys
 and non-zero for  Dy-alloys.

Finally, one can also measure the spin magnetic susceptibility of the
 alloys at low temperatures. 
 It should be  constant $(g\mu_{\rm B}J)^2/(2w)$ per ion in the Ho-alloys.
 In the Dy-alloys, an in-plane Curie susceptibility $(g\mu)^2/(3T)\approx 
(25g^2\mu_{\rm B}^2)/(3T)$ per ion 
is expected due to the magnetic moments.
 Unfortunately, the paramagnetic effect 
will be strongly masked by the Meissner susceptibility $\chi_M=-1/(4\pi )$
due to superconducting currents. 
One can, however, observe a remarkable 
weakening of the Meissner susceptibility at the temperature $\sim 0.1K$ 
for concentrations of Dy of about 1\%. At slightly lower temperatures the
Dy-alloys must go over into a spin-glass state. No such transition can 
be observed in the Ho-alloys.

In conclusion, we have shown that
the well-known difference between the magnetic properties of
ions of integer or half-integer total magnetic moments $J$
due to the absence or presence of Kramers degeneracy is captured
in the quasi-classical limit of large $J$ by a Berry phase.
In the tetragonal environment studied here,
 integer spins 
display no finite magnetic moment in the groundstate,
 whereas the 
half-integer spins display a finite magnetic moment of half their
saturation value. The difference is due to Kramers degeneracy \cite{abrag}
 necessarily associated with half-integer spins.
Any quasi-classical approach has to take the geometric Berry phase
into account in order to capture the essential difference between 
integer and half-integer spin.
The difference should be manifest in the spectrum of EPR, NMR and in the 
magnetic 
susceptiblity of dilute the alloys Lu$_{1-x}$Ho$_x$Ni$_2$B$_2$C and 
Lu$_{1-x}$Dy$_x$Ni$_2$B$_2$C.

Acknowledgements:
This work was partly supported by the NSF grant DMR-9705182 .
One of the authors (V.P.) thanks Professors T. Nattermann and J. Zittartz
for the hospitality extended to him during his stay at  Cologne University.


\begin{thebibliography}{99}
\bibitem{note1}
One can also discuss any of the excited states as long as these are
well-localized in the minima.
\end{thebibliography}
\end{document}